\documentclass[fleqn,10pt]{iopart}
\usepackage{graphicx}
\usepackage[uncertainty-mode=separate]{siunitx}
\usepackage{hyperref}
\usepackage[export]{adjustbox}
\usepackage{caption}
\usepackage{subcaption}
\DeclareSIUnit\angstrom{\text {Å}}
\begin{document}

\title{First Principles Validation of Energy Barriers in Ni$_{75}$Al$_{25}$}

\author{Adam Fisher$^{1,2}$, Julie B.\ Staunton$^3$, Huan Wu$^4$ and Peter Brommer$^2$}
\address{$^1$ EPSRC Centre for Doctoral Training in Modelling of Heterogeneous Systems (HetSys), University of Warwick, Coventry CV4 7AL, UK}
\address{$^2$ Warwick Centre for Predictive Modelling, School of Engineering, University of Warwick, Coventry CV4 7AL, UK}
\address{$^3$ Department of Physics, University of Warwick, Coventry CV4 7AL, UK}
\address{$^4$ TWI Ltd, Granta Park, Great Abington, Cambridge, CB21 6AL, UK }
\eads{\mailto{adam.fisher@warwick.ac.uk}, \mailto{p.brommer@warwick.ac.uk}}

\begin{abstract}
Precipitates in Nickel-based superalloys form during heat treatment on a time scale inaccessible to direct molecular dynamics simulation, but could be studied using kinetic Monte Carlo (KMC).
This requires reliable values for the barrier energies separating distinct configurations over the trajectory of the system. 
In this study, we validate vacancy migration barriers found with the Activation-Relaxation Technique nouveau (ARTn) method in partially ordered Ni$_{75}$Al$_{25}$ with a monovacancy using published potentials for the atomic interactions against first-principles methods.
In a first step, we confirm that the ARTn barrier energies agree with those determined with the nudged elastic band (NEB) method.
As the number of atoms used in those calculations is too great for direct \emph{ab initio} calculations, we then cut the cell size to 255 atoms, thus controlling finite size effects.
We then use the plane-wave density functional theory (DFT) code CASTEP and its inbuilt NEB method in the smaller cells.
This provides us with a continuous validation chain from first principles to kinetic Monte Carlo simulations with interatomic potentials. 
We then evaluate the barrier energies of five further interatomic potentials with NEB, demonstrating that none yields these with sufficient reliability for KMC simulations, with some of them failing completely.
This is a first step towards quantifying the errors incurred in KMC simulations of precipitate formation and evolution.
\end{abstract}

\noindent{\it Keywords\/}: Interatomic potentials, Kinetic Monte Carlo, Density Functional Theory, Validation

\section{Introduction}

Superalloys are alloys that can be used at a high percentage of their melting point without losing all of their desirable properties \cite{sims1984}.
Some of these properties include high strength, long fatigue life, fracture toughness, creep and stress-rupture resistance.
They also resist corrosion and oxidation at operating temperatures that other metallic compounds have limited use.
Nickel-based superalloys are used in a number of areas of  manufacturing including nuclear reactors, space vehicles, chemical processing vessels, submarines, and heat exchanger tubing \cite{Akca2015}, with their main use being in turbine blades in aviation \cite{Mouritz2012}.

\begin{figure}[bh]
    \centering
    \includegraphics[width=0.4\linewidth]{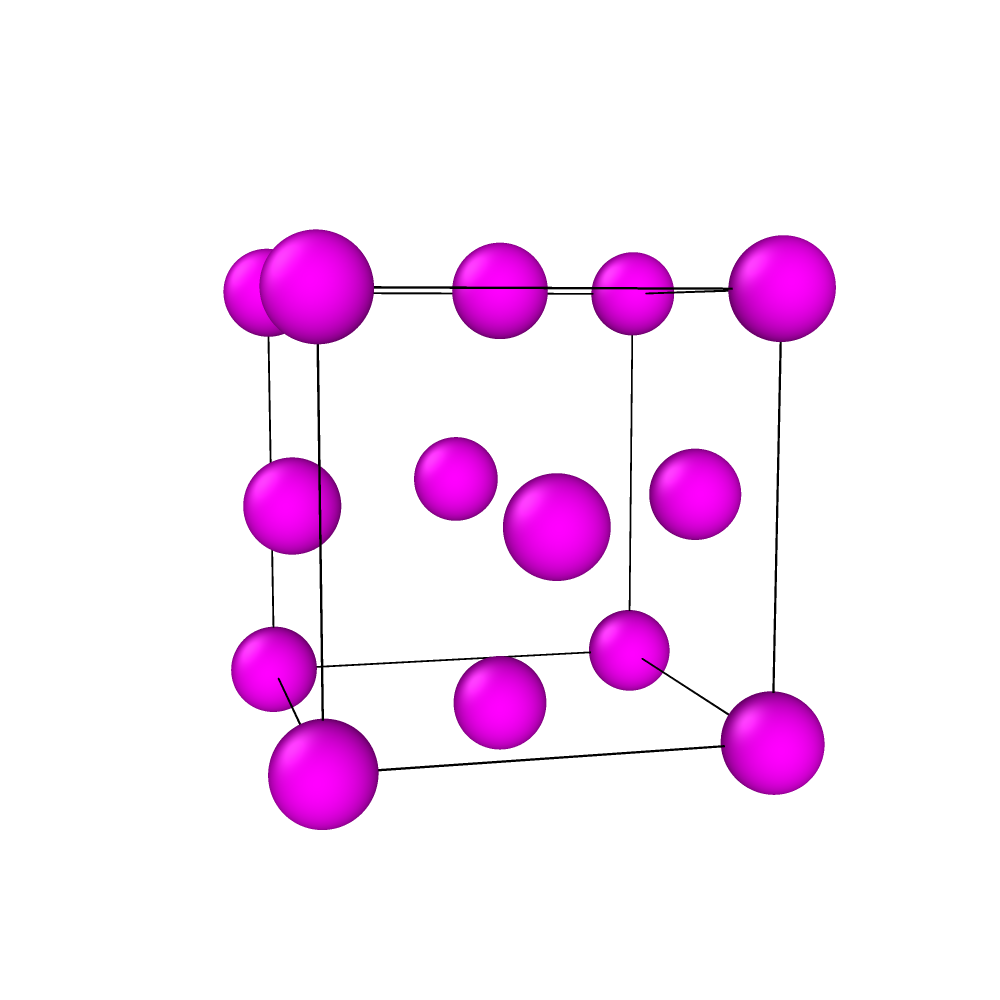}
    \includegraphics[width=0.4\linewidth]{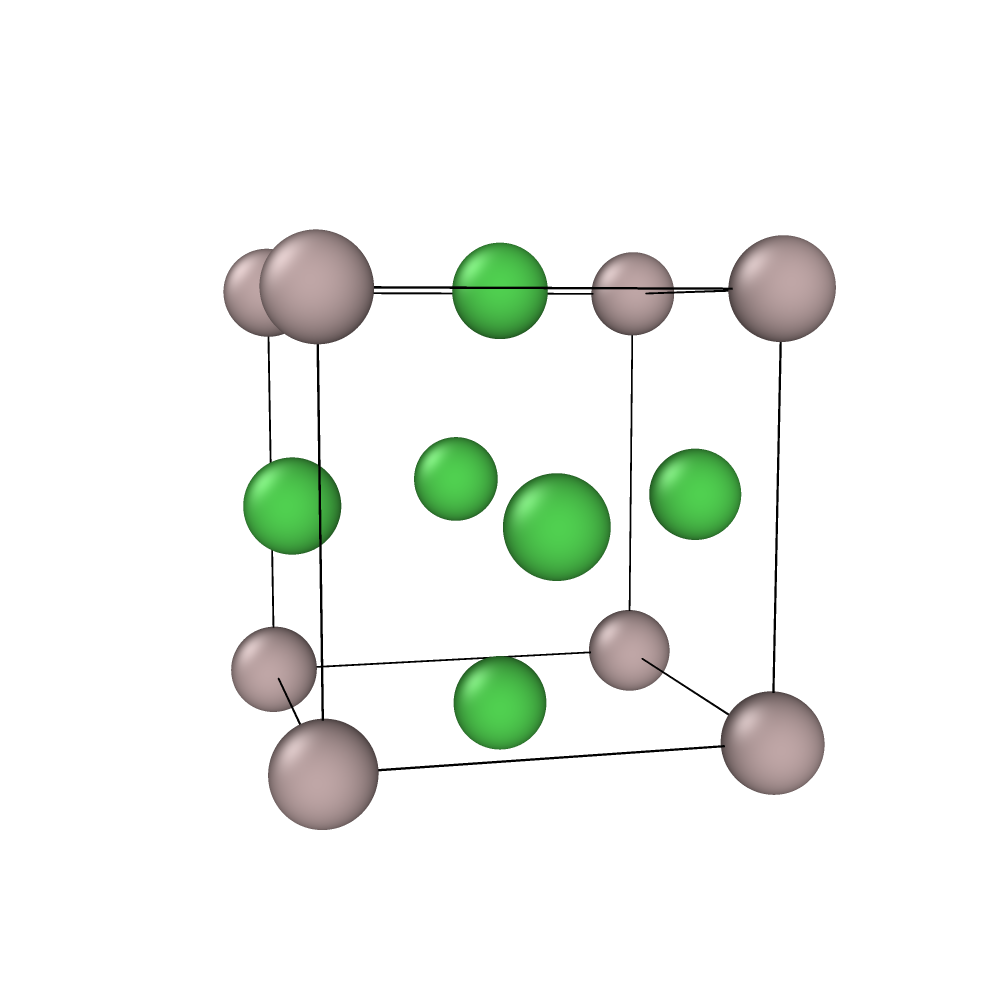}
    \caption{Left: $\gamma$ phase (pink atoms Ni or Al). Right: $\gamma{'}$ phase (green atoms Ni, grey atoms Al) \cite{Belan2016}. Made using ovito. \cite{Stukowski2010}}
    \label{fig:structure}
\end{figure}

In these applications, their desired properties are the resistance to fatigue and creep at temperatures of up to \qty{80}{\percent} of their melting temperature, coupled with their strength. 
These properties are determined by their microstructure, particularly how it affects material deformation behaviour \cite{panka_nathal_koss_1990}.
In nickel-based superalloys, there are two main phases:
Firstly, the $\gamma$ phase, a face-centred cubic (FCC) solid solution with no long-range order between the atomic species, forms the matrix that all other phases are embedded in.
Secondly, the $\gamma{'}$ phase, an ordered $L1_2$ structure (Fig.\ \ref{fig:structure}) with atoms sitting on FCC lattice sites, is usually either formed as Ni$_3$Al or Ni$_{3}$Ti. 
Another phase is the $\gamma^{''}$ phase which increases strength and forms a $D0_{22}$ ordered body-centred tetragonal lattice made of Ni$_{3}$Nb. This phase, however, decomposes at higher temperatures.
This leads to $\delta$ phases which are not in themselves weak but are not coherent with $\gamma$ phase, thus reducing strength of the overall alloy\cite{Belan2016}.
In addition, there are a number of topologically close packed (TCP) phases that form from other precipitants. 
These include $\sigma , \mu$ and Laves phases.
Research has shown that the interaction between $\gamma$ and $\gamma{'} $ phases also helps to increase superlattice intrinsic stacking fault energies which contributes to the creep resistance of these alloys \cite{Breidi2018}. 

The $\gamma{'}$ phase precipitates out of $\gamma$ during solidification and heat treatment; the size of the precipitates depends on the how the alloy has been treated \cite{Alabbad2019,Papadaki2018,Wu2019}.
The formation and evolution of the microstructure during this process is at the core an atomistic-scale process: The precipitation and growth of one phase in the other requires the reordering of atoms to occur and this reordering is governed by the kinetics of the atoms.
In absence of an atomistic understanding of these processes, highly prescriptive casting and treatment schedules have been developed for industrial applications, for example in turbine blades, to ensure consistent material properties.
However, these recipes cannot be applied to alternative manufacturing techniques such as additive manufacturing of alloy parts, where processing conditions are significantly different \cite{Frazier2014}.
In this case, a greater understanding of the atomic kinetics could support the creation of alloys with tightly controlled distribution of phases, which in turn might allow alloy design for their specific use conditions.

Unfortunately, the processes involved in the formation of microstructure cover timescales from femtoseconds (the time scale of thermal vibrations) up to hundreds of hours in heat treatment \cite{THOMAS2006}. 
Therefore, modelling how exactly these phases form, and by which processes, is a challenge.
Continuum scale approaches such as phase field methods \cite{Karma2001} rely on empirical laws to model the systems with limited links to bottom-up approaches.
These methods do not provide insight into atomic motion, and they do not explain why certain elements end up at the grain boundary. 
On the other hand, classical molecular dynamics (MD), which integrates the motion of atoms with timesteps on the order of femtoseconds, cannot access the time scales required. 
Kinetic Monte Carlo (KMC) \cite{Voter2007}, wherein the system is moved across energy barriers from one metastable state to the next, attempts to bridge the gap, as it can access longer time scales than molecular dynamics whilst showing both how atoms diffuse through the structure and how different phases start to form and grow. 
As such, simulated times of \SI{0.5}{\second} have been achieved using KMC methods \cite{Chatterjee2010}.

Kinetic Monte Carlo simulations require knowledge of the energy of a system given its atomic arrangement, both in metastable states and at the barriers separating them.
The most frequently used approach to achieve this is through use of interatomic potentials (IP).
These approximate the potential energy surface of a system using comparatively few parameters, thus making the energy cheap to calculate (on the order of a few \unit{\micro\second} per atom) \cite{Byggmästar2022}.
While IPs have successfully been used in the simulation of Ni-based superalloys to find diffusion mechanisms \cite{Duan2006,Duan2008}, calculate interface free energy \cite{Mishin2014}, study strengthening through microscale features \cite{Zhou2022} and analyse vacancy and dumbbell interstitial diffusion traps \cite{Ferasat2020}, their use for transition states and associated barrier energies required for KMC has not been demonstrated to date.
In investigating such use, it should be noted that KMC crucially relies on accurate energy barriers to find the correct kinetic evolution of the system. 
As such, it is prudent to assess IPs for use in KMC simulations based upon on their ability to correctly determine energy barriers.  
In this work, we propose, implement and demonstrate a validation protocol for the use of IPs in evaluating transition states and thus energy barriers.
In order to do so, we compare against first-principles methods to benchmark the accuracy of the IP used.
However, it would be computationally extremely costly to directly compare barriers found with open-ended barrier search methods, as used in KMC between IPs and first-principles methods.
Instead barriers are compared using the closed-ended nudged elastic band method in a cut-down simulation cell.
We apply this protocol to a range of published IPs. 

A summary of the underlying methods is provided in Sec.\ \ref{sec:Methods}: 
Fundamentals of KMC simulations in Sec.\ \ref{kmc}  and a description of the implementation in Sec.\ \ref{kart}.
An alternative barrier-finding method is outlined in Sec.\ \ref{NEB}, followed by the description of the first-principles method used (Sec.\ \ref{DFT}).
The selection of alternative IP is covered in Sec.\ \ref{OtherIPsMethods}.
The results, both for a reference IP and a wider selection of alternative IPs, are provided in Sec.\ \ref{sec:Results}. 
A discussion of these results follows in Sec.\ \ref{sec:Discussion}, and the main conclusions are provided in  Sec.\ \ref{sec:Conclusion}.

\section{Methods}\label{sec:Methods}

\subsection{Kinetic Monte Carlo}\label{kmc}
KMC is a method that studies the kinetics of a system over time as it evolves through a chain of metastable states, separated from each other by energy barriers \cite{Voter2007}.
KMC uses barrier heights to calculate the rate at which the system changes metastable state based upon transition state theory (TST).
TST examines the relative probability of finding a system at a boundary to finding a system at a minimum. 
In most KMC simulations, a harmonic approximation to transition state theory is used, known as harmonic transitional state theory (HTST).
In HTST, a rate is found using 
\begin{equation} \label{HTST}
    k^\text{HTST}=Ae^{(-E_\text{barrier}/k_BT)}
\end{equation}
where $k^\text{HTST}$ is the rate constant, $E_\text{barrier}$ is the barrier height, $T$ is the temperature. 
$A$ is a prefactor that is an approximation to  
\begin{equation} \label{prefactor}
    A = \frac{\prod\limits_i^{3N}\nu_i^\text{min}}{\prod\limits^{3N-1}_i\nu_i^\text{sad}}
\end{equation} 
where $\nu_i^\text{min}$ are the $3N$ normal mode frequencies at the minimum and $\nu_i^\text{sad}$ are the $3N-1$ non-imaginary normal mode frequencies at the saddle point \cite{Voter2007}.
As most prefactors in dense metallic systems are in the range of $\qtyrange{e12}{e13}{\per\second}$ \cite{Grabowski2018,Goswami2014} the prefactor is approximated as independent of the saddle point and thus does not need recalculating for each barrier. 
Once the rates $k_i$ of all barriers separating the current metastable state from neighbouring states have been calculated, a barrier to cross is selected at random weighted with their contribution to the total rate $k=\sum{k_i}$. 
The simulation clock is then advanced by an increment drawn from an exponential distribution with mean $\frac1k$, corresponding to the time to the first escape from the metastable state.
The system is then moved to the metastable state on the other side of the barrier and the process is resumed from there.
The process of moving from an initial state across a barrier to a final state is frequently called an event, characterised by the barrier energy $E_\text{barrier}$ and the energies of initial and final metastable state.

\subsection{Kinetic Activation Relaxation Technique}\label{kart}
In our study, we used the KMC implementation kinetic Activation Relaxation Technique (kART) \cite{El-Mellouhi2008,Béland2011}.
This is an off-lattice, self-learning KMC code that uses ART nouveau \cite{Barkema1996,Malek2000,Machado-Charry2011} to find the barriers connecting a state to neighbouring metastable states in a three-stage process. 
First, an atom and its neighbours are displaced along a random direction and the eigenvalues of the Hessian matrix are monitored.
When one of them is sufficiently negative, the system is deemed to have left the harmonic well. 
Next, the system is pushed in this direction while energy is minimised in the perpendicular hyper-plane, until we find the saddle point. 
This gives us the barrier energy. 
The final step is to push the system across the saddle and relax to the new minimum. 
This process is repeated until all atoms in the system have been sufficiently sampled for potential events, out of which one is then selected according to KMC rules.
KART uses a topological classification scheme to characterise the local environment of each atom, assuming that atoms with the same local environment structure will have access to the same events.
It then catalogues events by the topological identifier of the atom with the furthest displacement at the initial state, saddle point and final state and stores them in a database.
This has two main benefits: 
Firstly, it allows kART to identify if an event found by ART nouveau has been seen before. 
Secondly, if a certain local atomic environment reappears at a later step, the events accessible to an atom in this environment can be restored from the catalogue.
At that point, only the saddle points are re-converged, which dramatically reduces the computational cost.

\subsection{KMC simulation}
\begin{figure}[h!]
    \centering
	\includegraphics[width=0.7\linewidth]{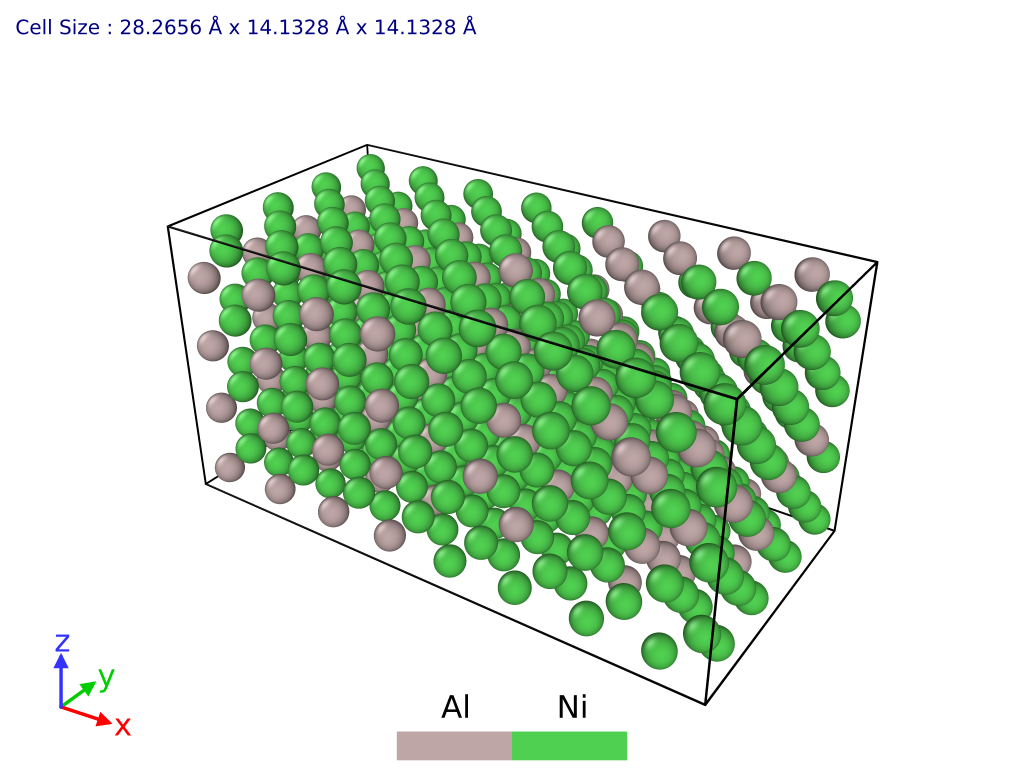}
	\caption{50\% $\gamma{'}$ (left) and 50\% solid solution of 75\% Ni and 25\% Al (right), with periodic boundary conditions. Visualised using ovito \cite{Stukowski2010}.}
	\label{fig:cell_setup}
\end{figure}

The simulation was set up as an $8\times 4\times 4$ supercell of the cubic unit cell.
Half of the cell was the ordered $L1_2$ $\gamma{'}$ phase, while the other half was a Ni$_{75}$Al$_{25}$ solid solution $\gamma$ phase as shown in Fig.\ \ref{fig:cell_setup}.
This allowed for us to watch an interface between the two phases and to look for growth of a phase.
We evolved this system over 500 KMC steps at \SI{900}{\kelvin} using the Molecular Dynamics package LAMMPS \cite{Plimpton1995} as a force calculator.
LAMMPS was used as it gives access to the OpenKIM database \cite{Elliott2011} of IPs and for this simulation we used the OpenKIM implementation of the Pun-Mishin 2009 (PM09) potential \cite{Elliott2018,Mishin2018, Pun2009} for NiAl.

The PM09 potential is an Embedded Atom Method (EAM) \cite{Daw1984} potential which is extensively used in atomistic simulations in the NiAl system, with over 300 citations at the current time. 
The authors describe it as ``highly transferable''  \cite{Pun2009}.
The Ni--Al cross terms have been fit to provide a good description of physical properties of L12-Ni$_3$Al and B2-NiAl. 
For these reasons, we used the PM09 potential as the baseline IP in this work. 
 
\subsection{Barrier calculations with the Nudged Elastic Band method}\label{NEB}
To avoid costly first-principles KMC simulations, we first compared the barrier energies found with kART to those computed with the nudged elastic band (NEB) method \cite{Jonsson1998}, which finds the minimum energy pathway of reaction coordinates between initial and final state. 
NEB is thus a closed-ended barrier finding method which unlike ART nouveau requires \emph{a priori} knowledge of the final state.
To determine the pathway, NEB creates a series of images between 2 potential wells; these images are connected by springs.
Forces on each image are calculated by finding the spring forces projected on to path between images and the atomic forces perpendicular to that path.
For our NEB calculations, we used LAMMPS \cite{Plimpton1995,Henkelman20009901,Henkelman20009978,Nakano2008,Maras2016} with 7 images connected by springs of strength \qty{1}{\electronvolt\per\angstrom\squared}.
Initial and final states were taken from kART.
The barrier energies found with NEB were then compared to those from the kART simulation to assure both methods found the same barriers.
At this stage, events that involve significant displacements of more than one atom (multi-atom events) were excluded, as they require larger cells to validate with first-principles methods.
Additionally, the barriers were unstable to decompose into multiple events involving consecutive motion of individual atoms. 
\subsection{Density Functional Theory simulations}\label{DFT}
Once we established that the barriers found by kART could be reproduced with NEB, we moved on to validating the barriers against a first-principles method, Density Functional Theory (DFT), to assess the accuracy of the IP used.
We used the DFT package CASTEP \cite{Clark2005} with the revised Perdew-Burke-Ernzerhof functional (rPBE) \cite{Hammer1999}.
Our settings for our DFT were a cut off of \SI{900}{\electronvolt}; an energy tolerance of \SI{2E-05}{\electronvolt\per atom}; a force tolerance of \SI{5E-02}{\electronvolt\per\angstrom}; a maximum displacement of \SI{1E-03}{\angstrom}; a maximum stress component of \SI{0.1}{\GPa} and $1$ k-point.
In a first step, we determined the DFT lattice parameter of disordered Ni$_{75}$Al$_{25}$.
To do this we took ten solid solution Ni$_{75}$Al$_{25}$ cells that had been randomly generated, we then relaxed them and averaged over the ten cells to get our lattice parameter.
We then used this lattice parameter for all further DFT calculations in order to recreate the pressure conditions that would be experienced in a bulk solid solution.
However our cells were still too large and, in order to validate the barriers, we first cut down our $8\times 4\times 4$ cells to $4\times 4\times 4$, to reduce computational cost of DFT.
We did this by centering the simulation box on the atom undergoing the largest displacement and cutting out a $4\times 4\times 4$ cell around that central atom.
Having cut down the cells, we repeated the LAMMPS NEB calculations to quantify the error incurred from cutting down the cells.

In a second step we calculated the barrier heights of events in the reduced $4\times 4\times 4$ cells using the NEB implementation in CASTEP.
For this we used 7 images and the ode12r preconditioner \cite{Makri2019}.
Given the computational cost of CASTEP NEB, we evaluated a subset of barriers found in the KMC simulation.
We prioritised pairs of events that had the same initial configuration but crossed different barriers and led to different final states.
Among those, we then sampled pairs of events at random.

Comparing barriers for events starting from the same initial configuration allows validation whether at least the relative ordering of barriers is maintained, even if the energies are off.
If relative barrier heights are maintained, a KMC simulation with an IP will only produce an incorrect timescale, whereas a change in barrier sequencing could lead to the KMC simulation exploring distinct pathways between first principles and IPs.

\subsection{Validation of Different IPs}\label{OtherIPsMethods}
Having set out a method for validating IPs for KMC simulations, we then explored the selection of existing IPs to validate.
We used the OpenKIM \cite{Elliott2011} database of IPs as a source.
At the time of our study, this database contained twelve potentials that could provide the interactions for nickel-aluminum systems:
eight embedded-atom method (EAM) \cite{Daw1984} potentials \cite{Pun2009,Angelo1995,Brommer2006,Farkas2020,Jacobsen1996,Mishin2004,Mishin2002,Pun2015} and four second-nearest neighbour modified EAM (2NN-MEAM) \cite{Baskes1992,Lee2000} potentials  \cite{Costa2007,Mahata2022,Kim2015,Kim2017}

Of those, two \cite{Jacobsen1996,Farkas2020} are designed to be used for more than just NiAl  and give up some accuracy for generality that is not needed in our system.
Two further potentials are not designed to capture the $\gamma{'}$ phase \cite{Brommer2006,Mishin2002}. 
Finally, both potentials by Kim \emph{et al.}\ \cite{Kim2015,Kim2017}  use \cite{Costa2007} for its NiAl interactions with additional elements, thus perform the same as \cite{Costa2007} in the present context as additional elements are not used.
This leaves, aside from the baseline PM09 potential \cite{Pun2009}, four further potentials to be validated for their suitability for barrier calculations: two EAM potentials (Angelo-Moody-Baskes [AMB]  \cite{Angelo1995,Baskes1995}, Mishin2004 [M04] \cite{Mishin2004}) and two 2NN-MEAM  potentials (Costa-Agren-Clavaguera [CAC] \cite{Costa2007}, Mahata-Mukhopadhyay-Asle-Zaeem [MMAZ] \cite{Mahata2022}.

\section{Results}\label{sec:Results}

\subsection{Validation of Pun-Mishin Potential}\label{PunMishin}
\begin{figure}[hbtp] 
    \centering
	\includegraphics[width=0.45\linewidth]{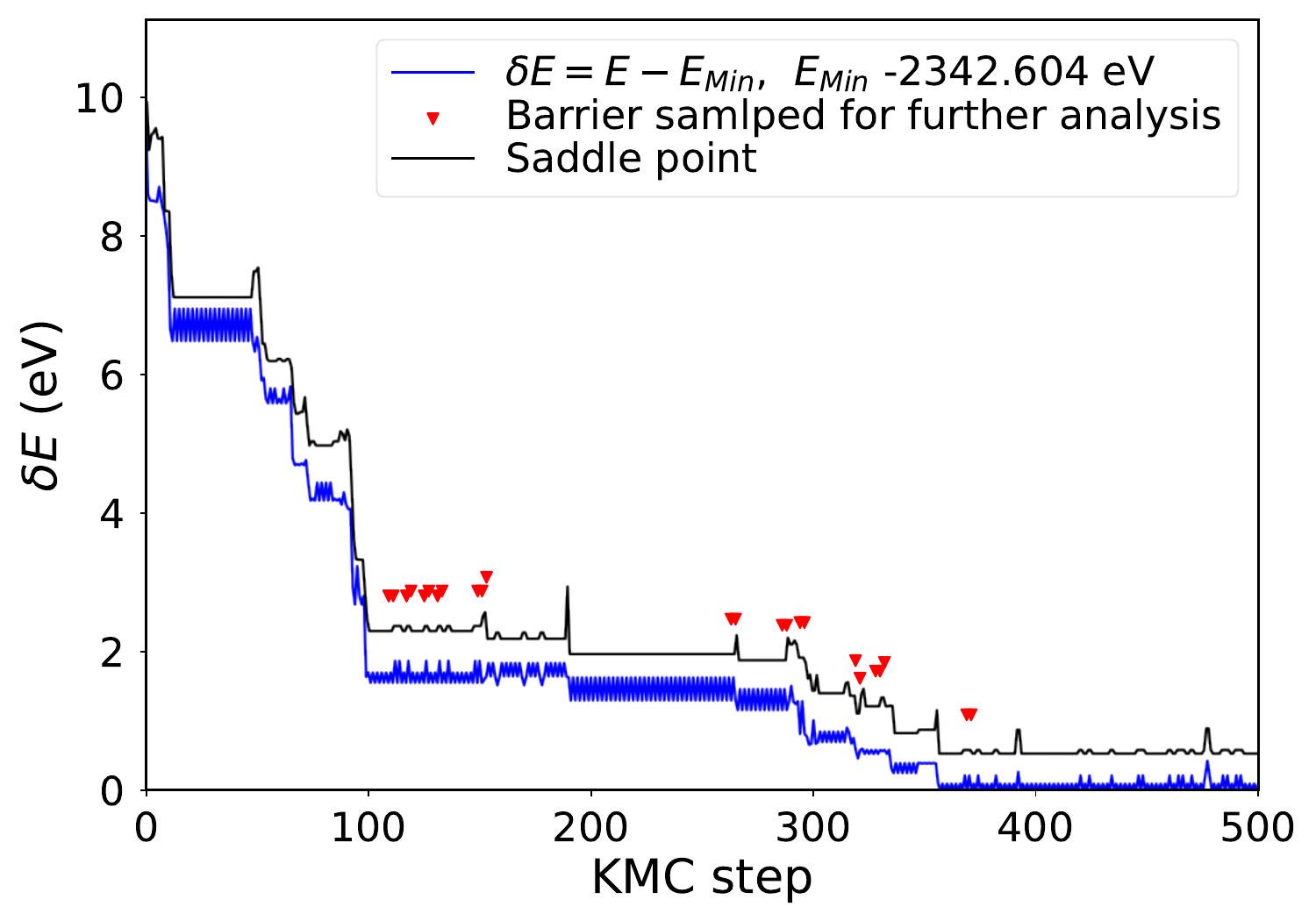}
	\caption{Energy of the simulation at the minima of each step (blue) and energy of the saddle point at each step (black). Red arrows point to the barriers used for validations against DFT}
	\label{fig:EnergyvsStep}
\end{figure}

In the kART simulation, over 500 events covering a simulated time of \qty{0.199}{\micro\second}, the system crossed $500$ barriers and reduced its potential energy by almost \qty{10}{\electronvolt}.
All events corresponded to a migration of the single vacancy inserted into the system.
The evolution of the system energy and the energy of the barriers crossed is shown in Fig.\ \ref{fig:EnergyvsStep}.
As can be seen from this figure, at various periods, the simulation crossed saddle points of very similar energies repeatedly, as shown by a horizontal line segment for the saddle point energies, e.g. between step 195 and 266, while the energy of the metastable states oscillates, indicating repeated back-and-forth crossing of the same barrier.
This is confirmed by an analysis of the kART trajectory, where the topological classification of initial state, saddle point configuration and final state is stored.
Only $111$ out of the $500$ events had a unique combination of initial, saddle and final state; the remaining 389 events show the same local environments at these three states, with individual events repeated up to 52 times.
Furthermore, 24 of the remaining events were found to be multi-atom events, i.e., a concerted motion of a chain of atoms, with the first one filling the vacancy and the vacancy moving to the site vacated by the final one.
These events are at best marginally stable against decomposition into multiple single-atom vacancy migration events.
This presented a challenge in analysing these barriers with NEB, thus they were excluded from further comparison.
This left us with only $87$ distinct single atom energy barriers that were considered for comparison.

\begin{figure}[hbtp]
    \centering
    \begin{subfigure}{0.45\linewidth}
        \includegraphics[width=\linewidth]{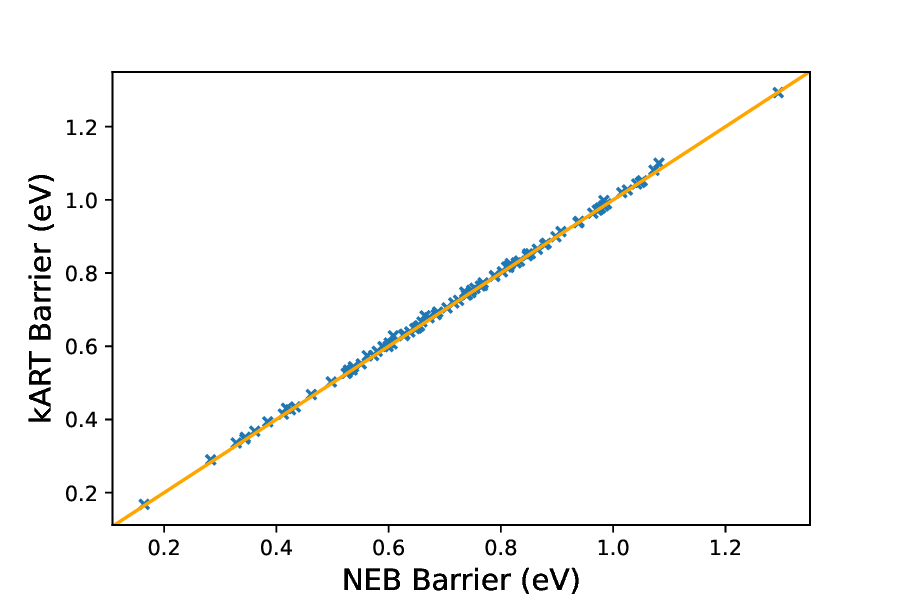}
        \subcaption{
        }
        \label{fig:kARTvsNEB-xy}
    \end{subfigure}
    \begin{subfigure}{0.45\linewidth}
        \includegraphics[width=\linewidth]{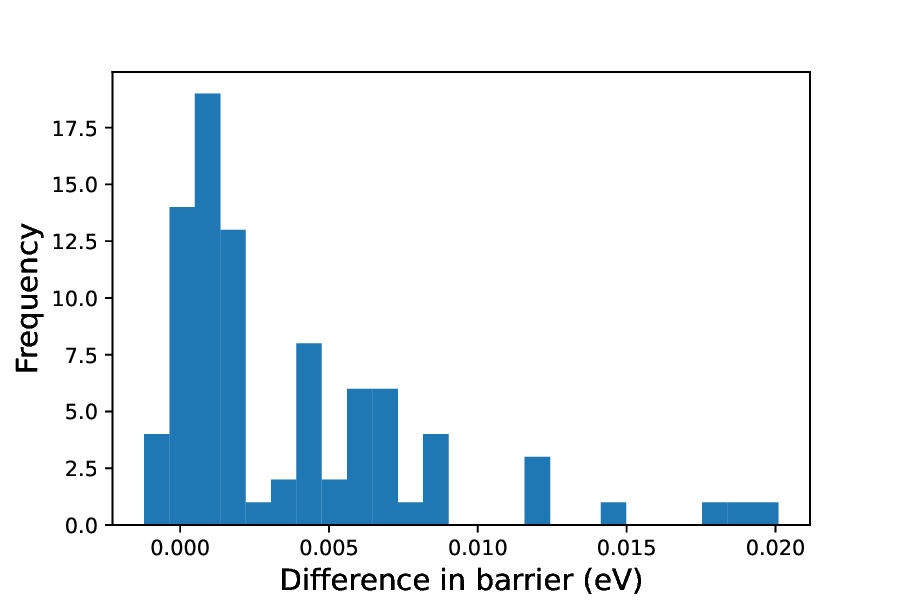}
        \subcaption{}
        \label{fig:kARTvsNEB-hist}
    \end{subfigure}
	\caption{Difference between distinct barriers values found using kART and NEB. (a) Barrier heights for kART and NEB. 
    (b) Histogram of the differences between kART and NEB, a negative values means NEB gets a higher value than kART.}
	\label{fig:meth_diff_distinct}
\end{figure}

When comparing barrier values obtained from kART and NEB, we see a very good agreement between the two different methods. 
It can be seen from Fig.\ \ref{fig:meth_diff_distinct} that at maximum there is a difference between methods of \SI{0.02}{\electronvolt} or maximum percentage error of \qty{3.2}{\percent} with most differing by much less than that. 
On average these two methods only differ by \SI{0.0037}{\electronvolt}; this level of agreement shows that the two methods employed produce consistent results.

\begin{figure}[hbtp]
    \centering
    \begin{subfigure}{0.45\linewidth}
        \includegraphics[width=\linewidth]{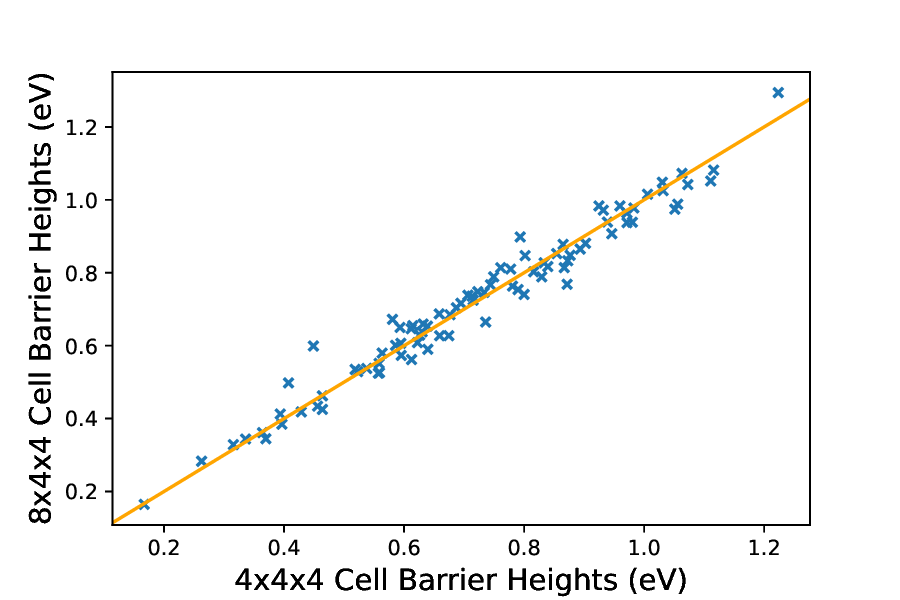}
        \subcaption{}
        \label{fig:CutvsUncut-xy}
    \end{subfigure}
    \begin{subfigure}{0.45\linewidth}
        \includegraphics[width=\linewidth]{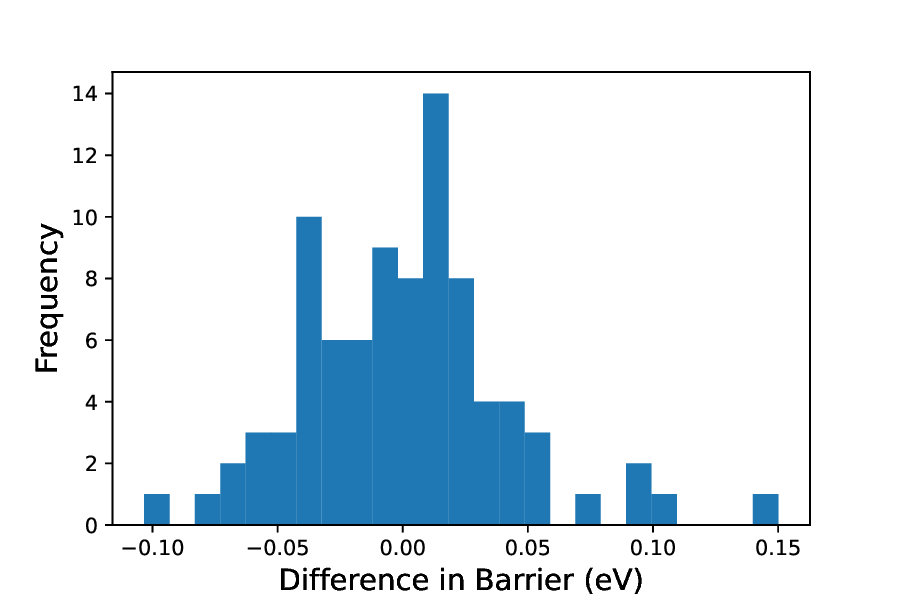}
        \subcaption{}
        \label{fig:CutvsUncut-hist}
    \end{subfigure}
	\caption{Difference between distinct barriers in cells of size $8\times 4\times 4$ and $4\times 4\times 4$. (a) Barrier heights in cells of size $8\times 4\times 4$ and $4\times 4\times 4$.  
    (b) Histogram of the differences between barriers in cells of size $8\times 4\times 4$ and $4\times 4\times 4$, a negative values means $4\times 4\times 4$ gets a higher value.}
	\label{fig:cell_diff_distinict}
\end{figure}

In the next step, we reduced the cell size by half to reduce DFT computational cost.
As shown in Fig.\ \ref{fig:cell_diff_distinict}, this finite-size effect changed the barriers by up to \SI{0.15}{\electronvolt}, with an average difference of \SI{0.031}{\electronvolt}.
This shows that the events we studied can be reasonably evaluated in the smaller simulation cell. 

\begin{figure}[hbtp]
    \centering
    \begin{subfigure}{0.45\linewidth}
        \includegraphics[width=\linewidth]{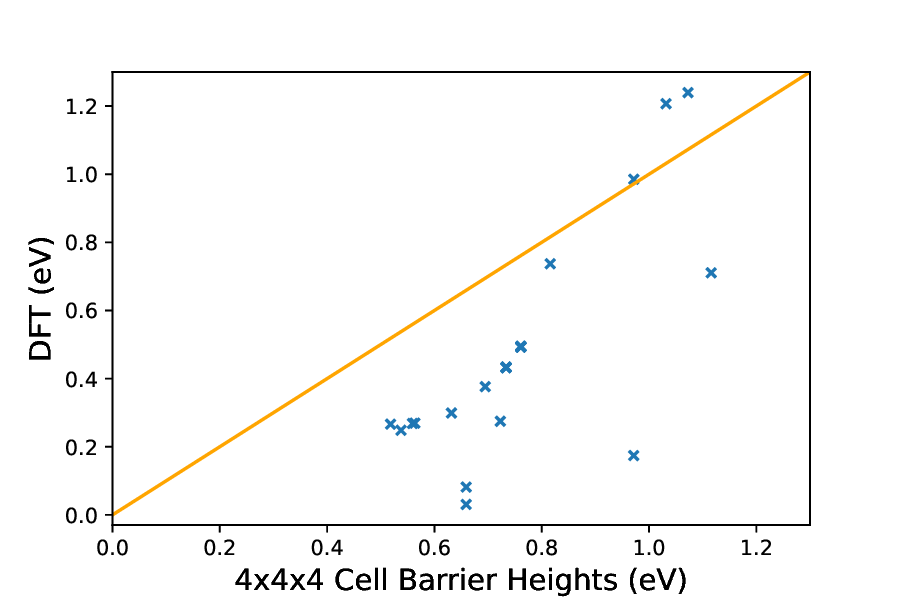}
        \caption{}
        \label{fig:PunVsDFT-xy}
    \end{subfigure}
    \begin{subfigure}{0.45\linewidth}
        \includegraphics[width=\linewidth]{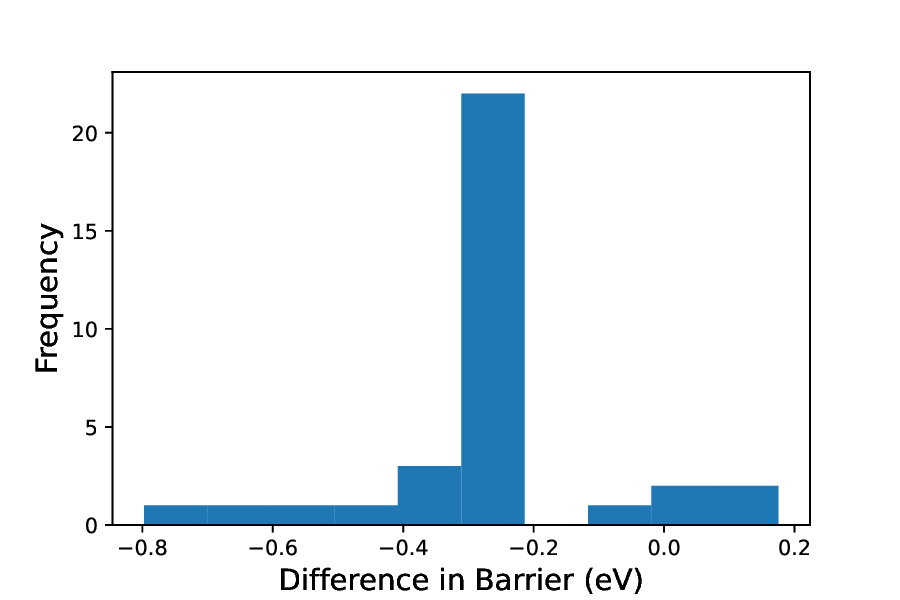}
        \caption{}
        \label{fig:PunVsDFT-hist}
    \end{subfigure}
    \caption{Difference in barrier heights DFT and Pun-Mishin. (a) Barrier heights for DFT and Pun-Mishin. 
    Histogram  of the differences between DFT and Pun-Mishin. Negative values are overestimations of barrier value compared to DFT.}
	\label{fig:DFT_diff}
\end{figure}

There were $32$ pairs of events identified and $13$ were selected at random to used.
Looking at the differences between the effective potential and DFT, PM09 overestimated the barriers by \SI{0.29(21)}{\electronvolt} as shown in Fig.\ \ref{fig:PunVsDFT-hist}. 

\begin{figure}[hbtp]
    \centering
	\includegraphics[width=\linewidth]{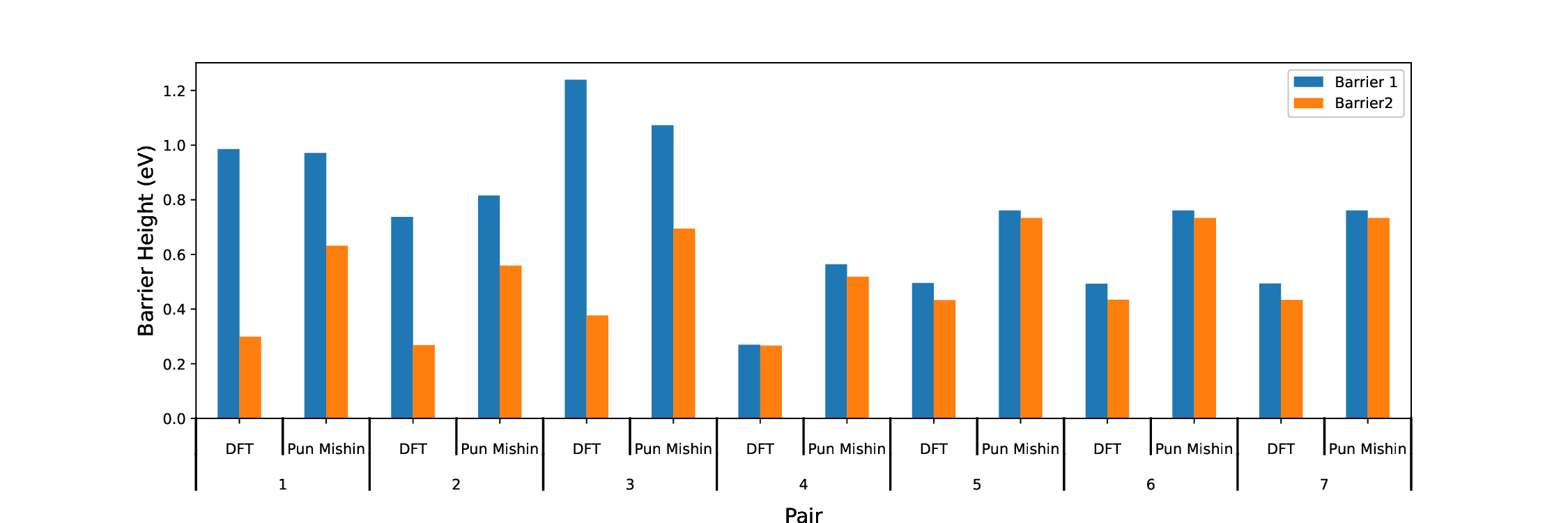}
    \includegraphics[width=\linewidth]{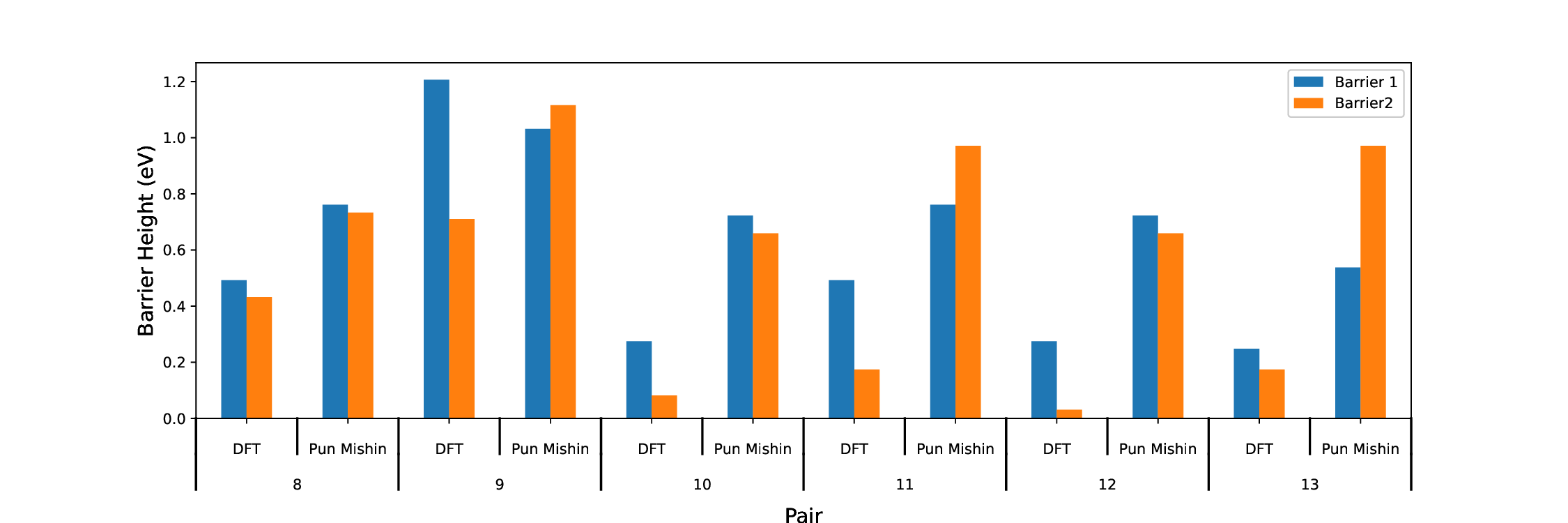}
	\caption{How pairs of barrier heights change from PM09 to DFT. Each barrier in a pair shares the same initial state.}
	\label{fig:Potential to DFT}
\end{figure}

If the overestimation was constant for all barriers, we would still see the correct kinetics in KMC but altered time scales, as constant offset in the barrier heights could be accounted for in a scaled prefactor in Eq.\ \ref{HTST}.
With the overestimation, timescales would be increased due to higher barriers meaning slower rates and a lower total rate for a KMC step.
However the distribution of EAM errors is wide, and for 3 of the starting states, the order of a pair of barriers switches, with the effective potential and DFT disagreeing on which one has the lower barrier height. 
This can be seen for pairs $9,11$ and $13$ on Fig.\ \ref{fig:Potential to DFT}.

This means using this potential could give us different kinetics, as what once was the most probable barrier is now less probable than another. 
This could lead the system to evolve through a completely different kinetic chain and result in completely unphyscial kinetics, although this is not guaranteed. 

As an illustration, the rates for event pair $9$ (Fig.\ \ref{fig:Potential to DFT}) are evaluated.
Assuming a constant prefactor of \qty{1e13}{\per\second} and a temperature of \SI{900}{\kelvin}, Eq.\ (\ref{HTST}) gives barrier 1 a rate of \SI{5.35e6}{\per\second} and barrier 2 \SI{1.71e7}{\per\second}  when calculated with PM09, whereas for DFT, the rates are \SI{1.06e9}{\per\second} and \SI{1.67e6}{\per\second}, respectively.
This leads to barrier 2 being selected with roughly triple the probability of barrier 1 for the interaction potential, while with DFT, barrier 1 is over 1000 times more probable to be selected than barrier 2.
\subsection{Validation of Other Interatomic Potentials}\label{OtherIPsResults}

When comparing other IPs to DFT, they performed the same or slightly worse on average than the PM09 IP. 
\subsubsection{M04}
\paragraph{}
The Mishin04 \cite{Mishin2004} (M04) potential had most of the barriers less than \SI{0.6}{\electronvolt} difference with all barriers tested under \SI{1.0}{\electronvolt} difference. 
This potential behaved similarly to PM09 in that vast majority of barriers sat in a narrow range of difference (\SI{0.5}{\electronvolt} and \SI{0.6}{\electronvolt} overestimate) with the rest spread over a much larger range of differences.
This high deviation from the average difference means this potential would be unsuitable for use in a KMC.
\begin{figure}[h!]
    \centering
    \begin{subfigure}{0.45\linewidth}
        \includegraphics[width=\linewidth]{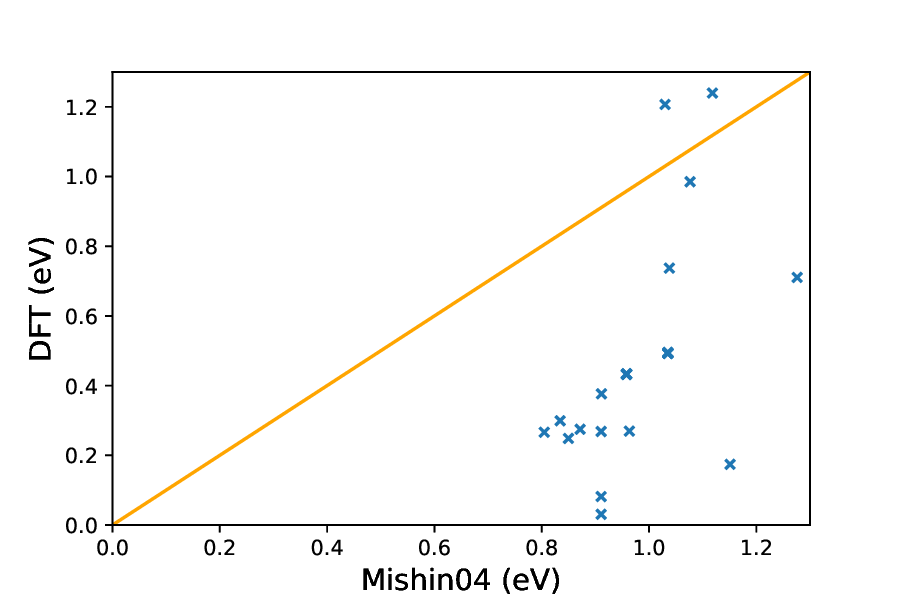}
        \caption{}
        \label{fig:MishinVsDFT-xy}
    \end{subfigure}
    \begin{subfigure}{0.45\linewidth}
      	\includegraphics[width=\linewidth]{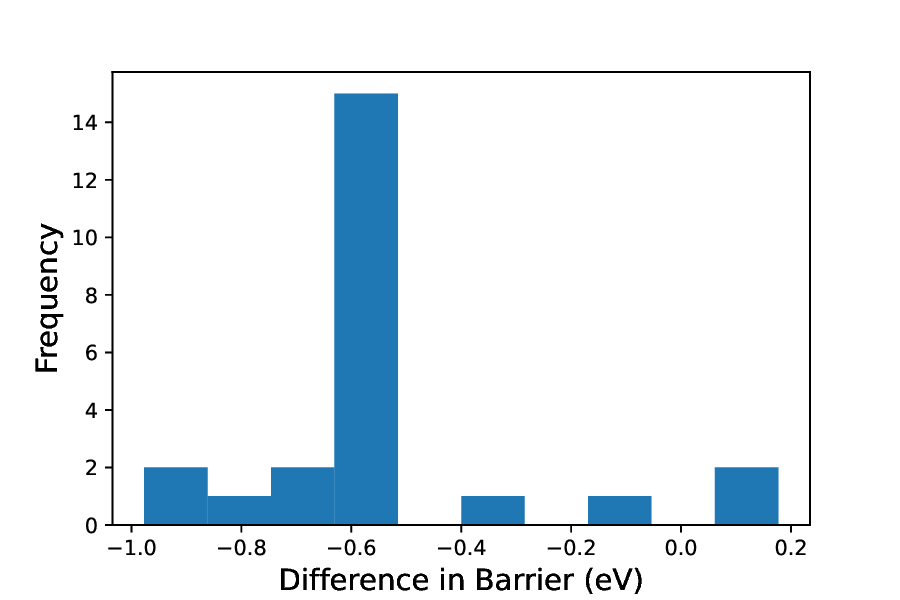}  
       \caption{}
       \label{fig:MishinVsDFT-hist}
    \end{subfigure}
	\caption{Barrier heights calculated with DFT and Mishin04 (M04) IP. (a) Barrier heights for DFT and M04.  
    (b) Histogram of the differences bewteen DFT and M04. Negative values are over estimations of barrier value compared to DFT.}
	\label{fig:Mishin04}
\end{figure}

\subsubsection{AMB}
\paragraph{}
\begin{figure}[h!]
    \centering
    \begin{subfigure}{0.45\linewidth}
        \includegraphics[width=\linewidth]{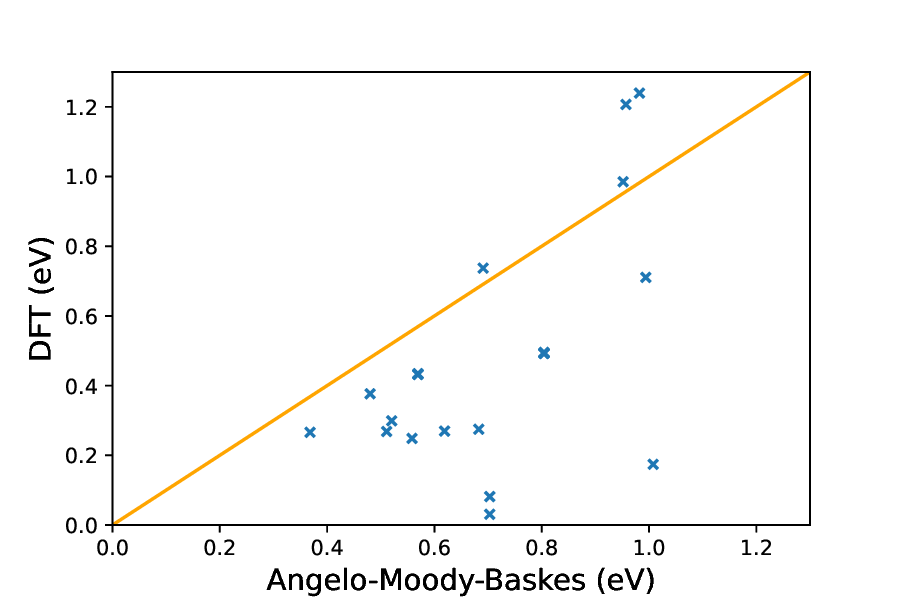}
        \caption{}
        \label{fig:AngVsDFT-xy}
    \end{subfigure}
    \begin{subfigure}{0.45\linewidth}
      	\includegraphics[width=\linewidth]{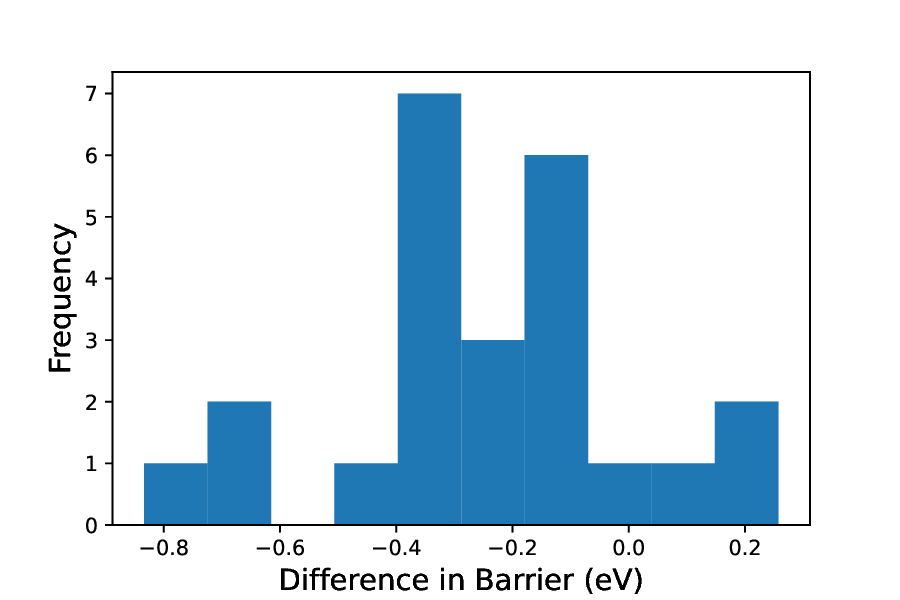}  
       \caption{}
       \label{fig:AngVsDFT-hist}
    \end{subfigure}
	\caption{Barrier heights calculated with DFT and Angelo-Moody-Baskes (AMB) IP. 
    (a) Barrier heights for DFT and AMB.  
    (b) Histogram of the differences bewteen DFT and AMB. Negative values are overestimations of barrier value compared to DFT.}
	\label{fig:Angelo}
\end{figure}

The Angelo-Moody-Baskes \cite{Angelo1995} (AMB) potential had most of the barriers less than \SI{0.4}{\electronvolt} difference but covered a range up to \SI{0.8}{\electronvolt}.
However, when comparing the distribution of differences between AMB and PM09 it is clear to see when comparing Fig.\ \ref{fig:AngVsDFT-xy} and Fig.\ \ref{fig:PunVsDFT-xy} that PM09 got closer to the DFT values more often.
AMB error in barriers was less consistent, as can be seen from comparing Fig.\ \ref{fig:AngVsDFT-hist} and Fig.\ \ref{fig:PunVsDFT-hist}.
PM09 potential had an overestimation of between \SI{0.2}{\electronvolt} and \SI{0.3}{\electronvolt} for a large majority of the barriers sampled whereas AMB had a wider distribution of values.
This wider distribution and larger error on average when compared to the already ruled out PM09 potential means it too is unsuitable for our purpose.

\subsubsection{MMAZ} 
\paragraph{}
\begin{figure}[h!]
    \centering
    \begin{subfigure}{0.45\linewidth}
                \centering
                \includegraphics[width=\linewidth]{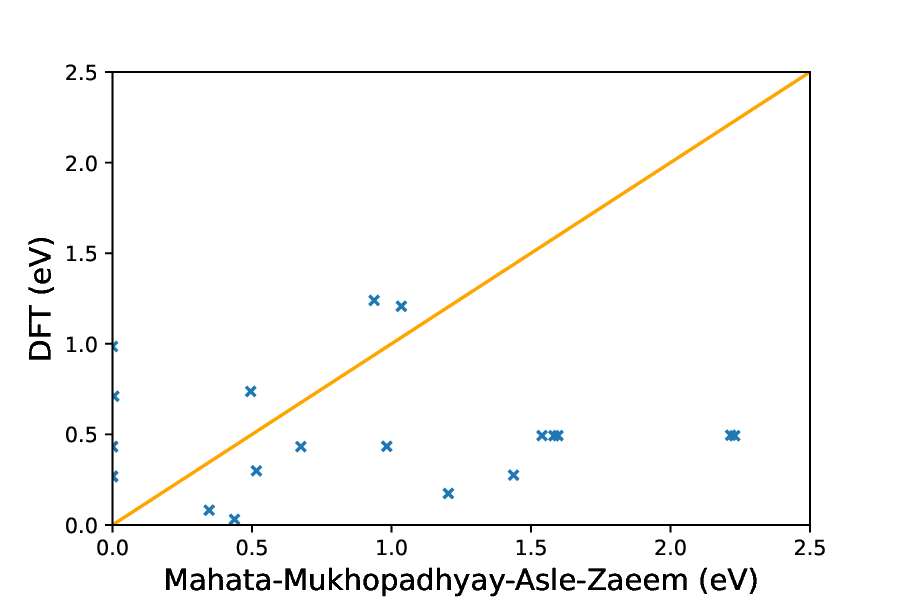}
                \caption{}
                \label{fig:MMAZ-xy}
    \end{subfigure}
    \begin{subfigure}{0.45\linewidth}
                \centering
                \includegraphics[width=\linewidth]{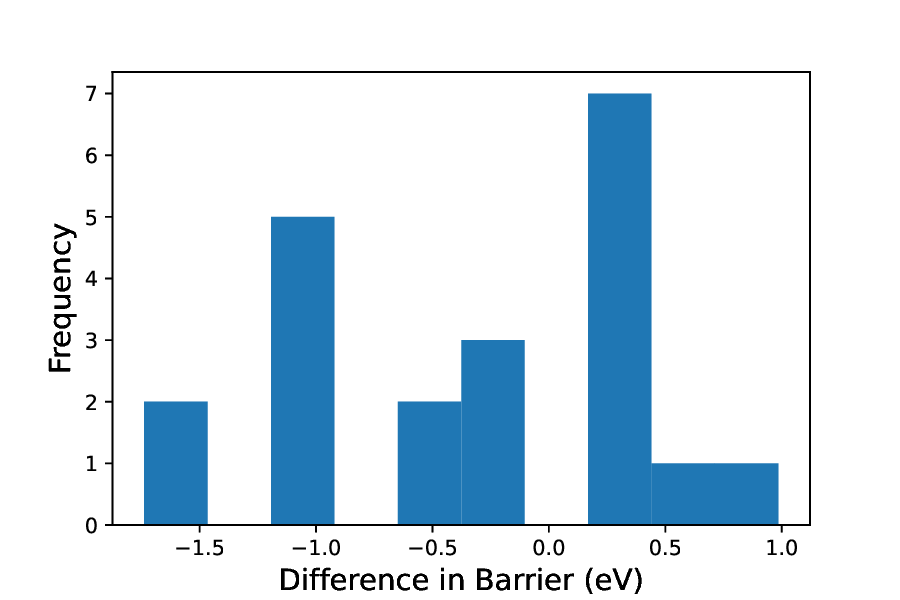}
                \caption{}
                \label{fig:MMAZ-hist}
    \end{subfigure}
	\caption{Barrier heights calculated with DFT and Mahata-Mukhopadhyay-Asle-Zaeem (MMAZ) IP. (a) Barrier heights for DFT and MMAZ. 
    (b) Histogram  of the differences between DFT and MMAZ. Negative values are overestimations of barrier value compared to DFT.}
	\label{fig:MMMAZ}
\end{figure}

Looking at the Mahata-Mukhopadhyay-Asle-Zaeem \cite{Mahata2022} (MMAZ) potential, it must be noted that the potential failed to recreate three of the barriers tested.
For the barriers that it recreated, most of the barriers had less than \SI{0.5}{\electronvolt} difference to DFT, however there were some barriers with differences up to \SI{1.7}{\electronvolt}. 
The combination of inability to reproduce some of the barriers and large range in differences means this potential could produce kinetics greatly different from that of DFT.
Therefore, this potential is not suitable for our purpose.

\subsubsection{CAC}
\paragraph{}
The Costa-Agren-Clavaguera \cite{Costa2007} (CAC) potential performs on average worse than AMB and PM09, with most differences less than \SI{0.4}{\electronvolt} and a larger range of differences of over \SI{1.2}{\electronvolt} between its barrier values and that of DFT.
This range in differences from DFT would make it unsuitable for use in a KMC.

\begin{figure}[h!]
    \centering
    \begin{subfigure}{0.45\linewidth}
        \includegraphics[width=\linewidth]{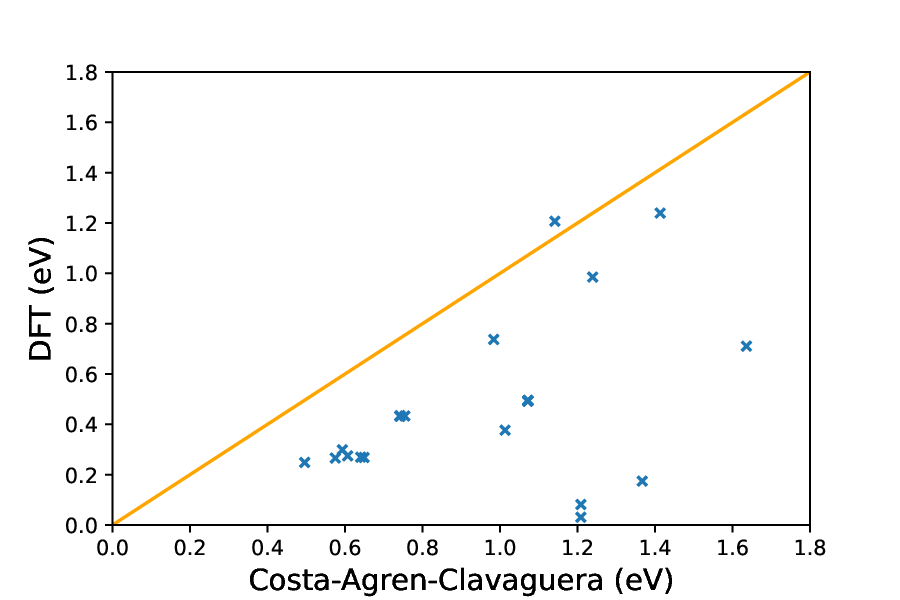}
        \caption{  
        }
        \label{fig:CosvsDFT-xy}
    \end{subfigure}
    \begin{subfigure}{0.45\linewidth}
        \includegraphics[width=\linewidth]{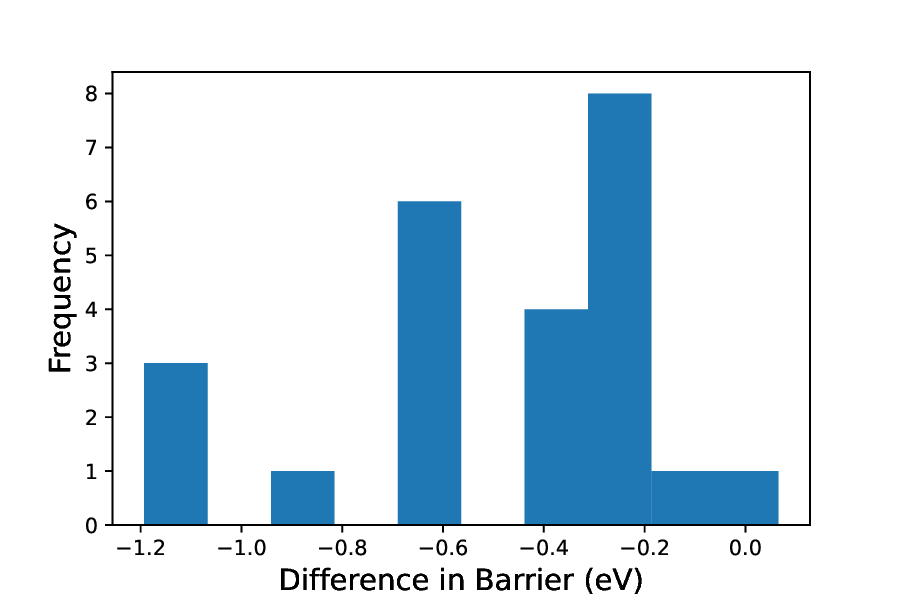}
        \caption{}
        \label{fig:CosvsDFT-hist}
    \end{subfigure}
	\caption{Difference in barrier heights DFT and  Barrier heights calculated with DFT and Costa-Agren-Clavaguera (CAC) IP. (a) Barrier heights for DFT and CAC. (b) Histogram of the differences between DFT and CAC. Negative values are overestimations of barrier value compared to DFT.}
	\label{fig:Costa}
\end{figure}

In summary, none of the potentials available in the OpenKIM database provide barrier energies with sufficient robustness to use in kinetic Monte Carlo simulations.
\section{Discussion}\label{sec:Discussion}
\subsection{KMC simulation}\label{sec:discusskmc}
While the main focus in this work is on the quality of energy barriers provided by a selection of IPs, the exemplar KMC simulation used to generate a selection of barriers already provides some insights in the challenges involved in these simulations.

Looking at Fig.\ \ref{fig:EnergyvsStep}, most of the evolution happens over relatively few KMC steps, separated by longer sequences of no progress (e.g.\ between step 195 and 266).
The evolution sequences are regularly preceded by the system crossing a different barrier compared to the barriers it uses in the ``calm'' periods, visible by a peak in the barrier energy line. 
This correlates with the \emph{replenish and relax} mechanism identified in recovery from radiation damage in silicon \cite{Béland2013}.
There it was found that key ``unlocking'' barriers need to be crossed before further relaxation is possible.
While the systems are quite different, there are also significant parallels:
In both cases, the systems are far from equilibrium (simulated radiation damage in \cite{Béland2013}, a random Ni$_{75}$Al$_{25}$ with a non-representative local ordering in this work), and the KMC simulation relaxes the systems towards the equilibrium state.
The existence of these key events reinforces the need for getting their barriers right: 
Not only do they determine the overall rate of relaxation to a much higher degree than the ``unproductive'' events preceding them.
An underestimated barrier energy for the latter will mainly cause computational issues, requiring the costly simulation of many events going nowhere.
A wrong key barrier also has the scope to derail the evolution of the system to another pathway, if alternative key barriers are then found to be more favourable.
It should be noted, that due to the procedure of how the pairs of barriers were identified in the EAM simulation, the system after crossing the first of a pair eventually returned to the same state to cross the other. 
This would indicate that the former is probably one of the ``unproductive'' events, and we are unable to say whether we actually compare two key event barriers.

There is also no certainty that the KMC simulation with IP finds all barriers that an extremely costly simulation with first-principles forces and energies would find.
An IP that would yield barrier energies consistent with DFT for the barriers it identifies would inspire some confidence that it does not miss crucial barriers, but none of the IPs under study here achieve this.

\subsection{IP Quality}\label{sec:ipdiscuss}
In this section, we discuss the performance of the IP discussed in this work and relate this to the original purpose of the IP. 
The performance of the PM09  potential is surprisingly good in a system ordering towards a $\gamma'$ structure:
It is only fitted on the energies of three ordered Ni$_3$Al structures and no disordered structures.
The rest of the information comes from different NiAl stoichiometries and a non-linear combination of pure Ni and pure Al potentials.
Given that newer machine learned interatomic potentials, such as Gaussian approximations potentials \cite{Bartok2010} or atomic cluster expansion potentials \cite{Drautz2019}, can be fit on thousands of structures, PM09 does not do too badly with comparably little information -- but obviously not quite good enough.
The results suggest that although the potential can relatively accurately calculate the correct energy for a number of the low energy configurations it struggles further away from these points. 
This then leads to a failure to reproduce barrier energy ordering in over \qty{20}{\percent} of cases, which does not give sufficient confidence that the PM09 EAM selects the right barrier in cases where it could actually lead to a different evolution of the system.
This restricts the suitability for using the PM09 IP in our KMC simulations.

The M04 potential \cite{Mishin2004} was shown to perform worse than PM09.
When looking at the fitting data of this potential, it focuses more on Ni$_3$Al at the expense of other NiAl phases compared to the later PM09 IP \cite{Pun2009}, with successfully reproducing experimental and first-principles quantities in this system. 
However, this apparently does not extend to transition state energies as shown here.

The AMB IP \cite{Baskes1995} was originally designed to study hydrogen embrittlement of Ni--Al alloys, based on an earlier Ni--H potential \cite{Angelo1995}, so its main purpose is not the simulation of NiAl structures.
Its Ni--Al interactions are fit to a combination of Ni$_3$Al and NiAl properties (lattice constants, sublimation energies plus Ni$_3$Al planar faults) similar to PM09 and it shows an accuracy approaching that of PM09.
Its main difference is the more prescribed functional form with fewer fitting parameters. 
However, the information used in the fitting process is unsurprisingly also insufficient to reproduce barrier energies. 

MMAZ's Ni-Al interactions are fit to properties of the B2 NiAl phase first, then the MEAM parameters are adjusted to reproduce the formation energies of other phases including the $\gamma'$ phase \cite{Mahata2022}. 
The elementary interactions are taken from \cite{Asadi2015}.
It is shown to reproduce lattice parameter, elastic constants, $\gamma$-Ni/$\gamma'$-Ni$_3$Al interface energy as well as Ni$_{75}$Al$_{25}$ liquidus temperature well, covering both low- and high-temperature properties.
However, it falls short in the description of barriers (including some barriers that it fails to find at all).
This seems to hint at the challenges of getting the saddle point energies right: some atoms are reasonably far from their equilibrium positions, but the overall order is still mostly consistent with a low temperature.

The CAC IP Ni-Al interaction \cite{Costa2007} was fitted ad-hoc to B2-NiAl and $\gamma'$-Ni$_3$Al lattice parameters, elastic constants and formation enthalpies to study $\gamma$-Ni/$\gamma'-$Ni$_3$Al interface energies, and combined with elementary MEAM IP from \cite{Lee2003}.
While this potential locates all barriers, its shortcomings in the barrier energies are evident.
So while MEAM potentials may give a better representation of alloy phase diagrams and related properties, the two MEAM potentials studied here struggle with vacancy migration energies in partially ordered systems.

\medskip

In summary, all tested interatomic potentials struggle in various degrees to provide a sufficient description of transition state energies.
The potentials were trained to describe a selection of physical properties in the Ni--Al alloy system, such as lattice parameters, formation energies or elastic constants.
However, this information does not cover transition states, where a limited number of atoms are displaced significantly from their equilibrium positions (typically up to half a nearest-neighbour distance).
In comparison, even for simulation of elastic constants, atoms are only displaced by a few percent of the nearest neighbour distance.
This shows that a good description of transition states requires a different approach to potential fitting -- they cannot be adequately extrapolated from reference data where all atoms are reasonably close to their equilibrium position.

\section{Conclusions}\label{sec:Conclusion}

In this work, we describe a tool chain to validate barrier heights found in kinetic Monte Carlo simulations calculated from interatomic potentials against density functional theory.
While we focus on the vacancy migration in the $\gamma/\gamma'$ NiAl superalloys, this method could be applied to other KMC simulations, which rely on the precision of barriers calculated with IP.
The tool chain is supported by a number of python scripts, which have been made publicly available.
Depending on the system under study, the step of embedding events in a suitably small unit cell for DFT simulation is the one requiring most manual intervention.

As part of this tool chain we found that kART as an open-ended barrier search method and NEB as a closed-ended one find consistent barrier heights for single-atom events. 
This is essential to the process, as computationally more expensive open-ended barrier searches are almost unfeasible with DFT methods and so a closed-ended method is required to allow this.

The validation of five IPs for the NiAl system showed that even though the PM09 interatomic potential fails to reproduce the correct order of barrier heights and thus may not be suitable for KMC simulations, it is the least bad of the interatomic potentials under study.
The other potentials studied show larger discrepancies compared to DFT, with AMB being almost comparable to PM09, while MMAZ fails to reproduce some of the barriers. 

This validation exercise demonstrates that energy barriers for vacancy diffusion cannot be assumed to be reproduced correctly.
This is particularly true for potentials fitted to data that does not include atoms with significant displacements from their equilibrium positions.
This agrees with work in other systems.
For example, vacancy migration barriers were identified as of crucial importance for long-time evolution, and they were made the focus of IP fitting in base-centred cubic iron \cite{Malerba2010}.
The validation protocol outlined here can provide a guide on how IPs can be tested for their ability to provide this information.
While the OpenKIM API \cite{Elliott2011} already provides tests for vacancy formation and migration in some elementary systems, a full coverage of energy barriers as they actually appear in a KMC simulation requires the more flexible approach outlined here.

The next question to answer now is how to improve the veracity of KMC simulations.
As no potential in the OpenKIM database provides barrier energies with sufficient confidence for KMC simulations, further studies of the long-term evolution of the $\gamma/\gamma'$ NiAl system would require a purpose-fitted interatomic potential.
For classical IPs, this could be achieved with force matching \cite{Ercolessi1994}, where DFT saddle point configurations and similar could be included in the reference data set used to fit the potential parameters.
Force matching implementations can handle the large datasets this approach would produce, as well as potentials with many parameters, including tabulated potential functions \cite{Brommer2015}.
On the other hand, a machine learned interatomic potential (MLIP) could offer the accuracy required by training on a sufficient set of data which would include these critical configurations.
Either option would be key to study atomistic processes on longer timescales, such as the evolution of the $\gamma/\gamma'$ microstructure.

\section*{Acknowledgments}

A F was supported by a studentship within the UK Engineering and Physical Sciences Research Council-supported Centre for Doctoral Training in Modelling of Heterogeneous Systems (HetSys), Grant No. EP/S022848/1.
This publication was made possible by the sponsorship and support of TWI’s Core Research Programme. The work was enabled through the National Structural Integrity Research Centre (NSIRC), a postgraduate engineering facility for industry-led research into structural integrity established and managed by TWI through a network of both national and international Universities. \footnote{\url{https://www.twi-global.com/crp}}
Calculations were performed using the Sulis Tier 2 HPC platform hosted by the Scientific Computing Research Technology Platform at the University of Warwick. 
Sulis is funded by EPSRC Grant EP/T022108/1 and the HPC Midlands+ consortium. 
We acknowledge the University of Warwick Scientific Computing Research Technology Platform for assisting the research described within this study.

\section*{Data availability and software}
Software and data for this paper are available at \cite{Fisher2024}. Current versions of the code can be downloaded from at \url{https://github.com/admfisher14/First-Principles-Validation-of-Energy-Barriers}

\bibliographystyle{iopart-num}
\section*{References}
\bibliography{References}

\end{document}